\documentclass{PoS}
\usepackage{braket}
\usepackage{mathtools}
\usepackage{qcircuit}
\usepackage{placeins}
\usepackage{subcaption}

\graphicspath{ {./figures/} }

\newcommand{\beq}{\begin{equation}}
\newcommand{\eeq}{\end{equation}}
\renewcommand{\>}{\rangle}

\title{ Lattice Gauge Theory for a Quantum Computer}

\ShortTitle{Quantum Computing LGT}

\author{\speaker{Richard C. Brower}\\
	Boston University, Boston, MA 02215, USA\\
				\email{email: brower@bu.edu}}

\author{David Berenstein\\
      University of California
Santa Barbara, CA 93106, USA\\
       \email{email: dberens@physics.ucsb.edu}}

\author{Hiroki Kawai\\
        Boston University, Boston, MA 02215, USA\\
       \email{email:  hirokik@bu.edu}}

\abstract{The quantum link~\cite{Brower:1997ha}   Hamiltonian 
was introduced two decades ago as an alternative to Wilson's Euclidean lattice
QCD  with  gauge fields represented by  bi-linear fermion/anti-fermion operators. 
When  generalized this new microscopic representation of lattice field theories is  referred as {\tt D-theory}\cite{Brower:2003vy}.
 Recast
as a Hamiltonian in Minkowski space for real
time evolution,  D-theory leads naturally to quantum Qubit  algorithms. Here to explore
digital quantum computing for gauge theories,  the simplest example of U(1) compact
QED  on triangular lattice is defined and gauge invariant kernels for the Suzuki-Trotter expansions are expressed as Qubit circuits capable of being  tested on the IBM-Q and other  existing Noisy Intermediate Scale Quantum (NISQ) hardware. This is a
 modest step in exploring the quantum  complexity of
D-theory   to guide  future applications to high energy physics and condensed matter quantum field theories. 
}

\FullConference{37th International Symposium on Lattice Field Theory - Lattice2019\\
		16-22 June 2019\\
		Wuhan, China}

\begin{document}

\section{Introduction}

Quantum Chromodynamics (QCD) remains a singularly difficult
computational problem. In spite of the success of Wilson's Euclidean lattice
QCD, there remains the notorious {\em sign problem} making predictions
of  real-time dynamics impossible, as well as severely limiting the
determination of parton distribution functions and thermodynamics at
non-zero chemical potentials needed by HEP experiments. On quantum
computers, these sign problems are solved in principle. One defines a
Hamiltonian operator for unitary evolution in continuous time.  Two
{\em discretizations} are required for any finite Qubit algorithm: {\bf i)}  Placing the Kogut Susskind Hamiltonian
\cite{Kogut:1974ag} on a finite volume lattice and replacing Euclidean time ($ i t$)  by Minkowski time ($t$) to guarantee
unitary real time quantum evolution.  {\bf
  ii.)}  The  less familiar step of 
replacing field variables by  a finite set of Qubits  per
lattice cell. The {\tt quantum link} representation of   Brower, Chandrasekar and Wiese~\cite{Brower:1997ha}  or {\tt D-theory} accomplishes this second, field space  {\em quantization}.

Since the  introduction of D-Theory in the context of
classical computing algorithms, it has continued to be  developed for  quantum computing of growing  number of interesting theories reviewed by
Uwe Wiese~\cite{Wiese:2014rla}.  This talk  will briefly present D-theory for
QCD and then consider a simple prototype  example of $U(1)$ gauge theory
on a 2d triangular lattice, which is capable of  testing small kernel  consistent with the hardware
constraints in the NISQ era. 

\section{Quantum link D-theory for QCD} 

\begin{figure}[h]
\begin{center}
  \includegraphics[width=.4\textwidth]{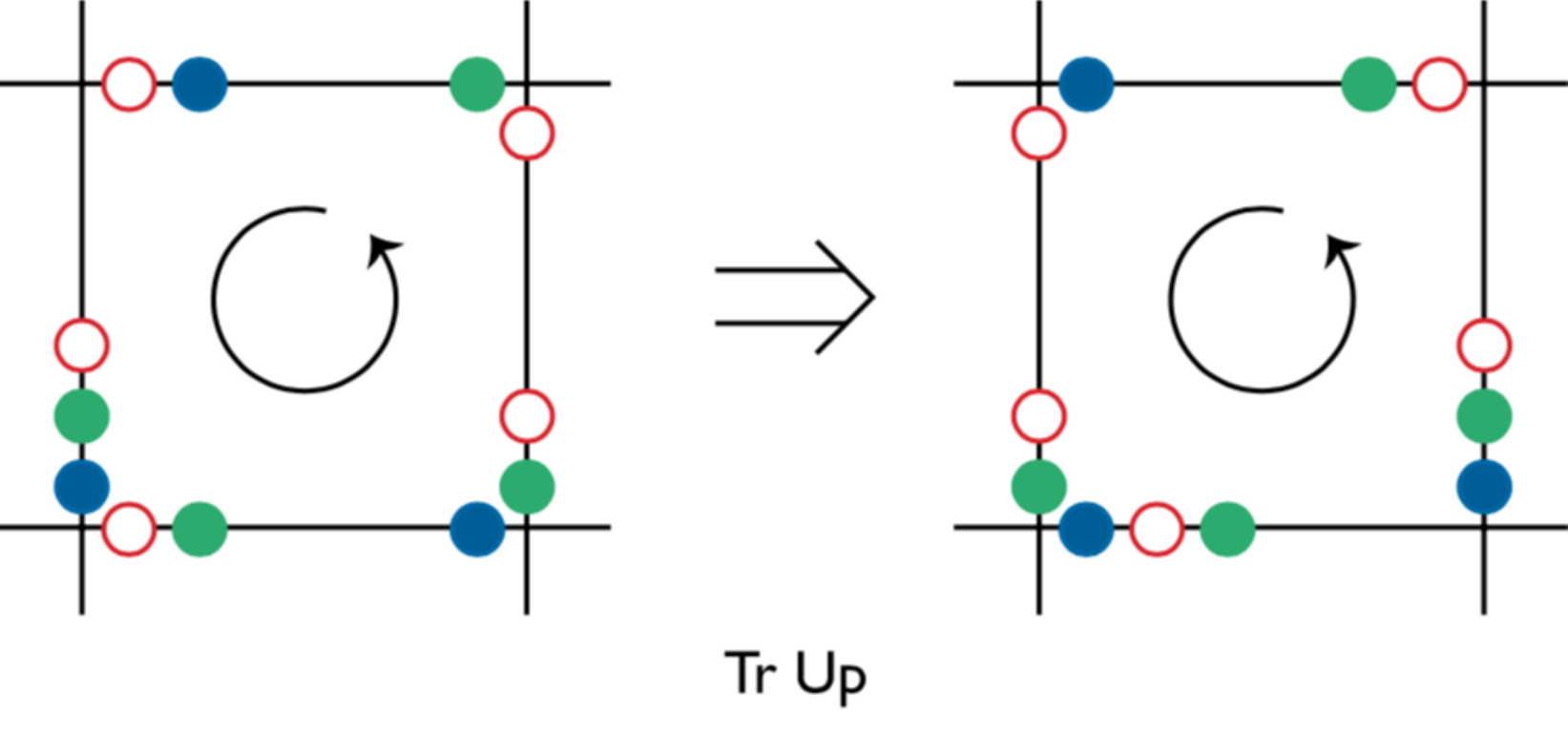}
  \includegraphics[width=.4\textwidth]{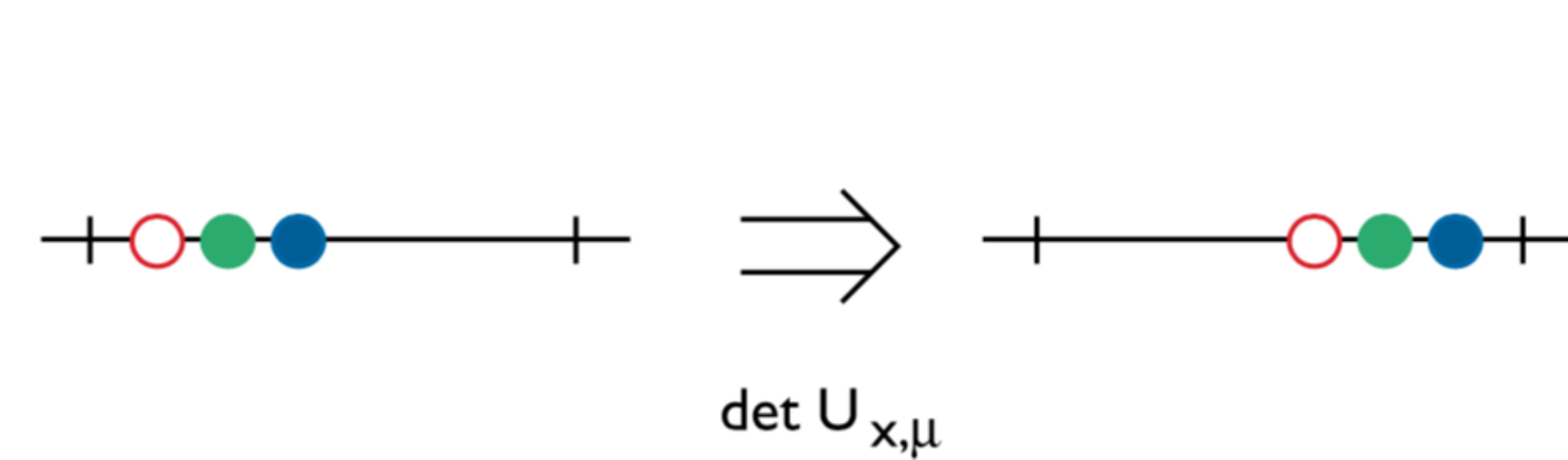}
\end{center}
\caption{\label{fig:QCDlink} QCD quantum link dynamics: On the left, the Hamiltonian induces 
a hopping of link fermion to rotate  various colors around a plaquette. On the right, 
by adding an extra determinant term the gauge group is broken from U(3) to SU(3) by  translating  
a {\em baryon-like}  object on the link.}
\end{figure}

In going from the Euclidean D-theory to the Minkowski Hamiltonian, what is
required, roughly speaking, is swapping the extra-dimension fermion with a
gauge-fixed temporal realtime axis. The result is a wide range of spin and
gauge field theories represented in the terminology of Bravyi-Kitaev~\cite{quant-ph/0003137} as {\em Fermionic Quantum
Computing} algorithms.  For example, for QCD the Hamiltonian is 
\beq 
\hat H = \frac{g^2}{2} \sum_{x,\mu} Tr[ \hat E_L^2(x,\mu) + \hat
E_R^2(x,\mu)] - \frac{1}{2 g^2}\sum_{x,\mu, \nu} Tr[\hat
U_{\mu\nu}(x)] + \mbox{Quarks}
\label{eq:Hhat}
\eeq
The key to the construction is a  faithful preservation of the symplectic gauge
algebra by introducing on each link ($\<x,x+\mu\>$) operators, 
\beq
\begin{split}
\hat  E_L(x,\mu) &= a^{i\dag }_{\<x,x+\mu\>} a^j_{\<x,x+\mu\>}\quad ,
\quad   \hat E_R(x,\mu) =    b^{i\dag}_{\<x,x+\mu\>} b^j_{\<x,x+\mu\>}\; ,
\; \\
 \hat U^{ij\dag}(x,\mu) &=  \hat  a^{i\dag}_{\<x,x+\mu\>}
 b^j_{\<x,x+\mu\>} \quad , \quad 
U^{ij}(x,\mu) =   b^{i\dag}_{\<x,x+\mu\>} a^j_{\<x,x+\mu\>}  \; ,
\end{split}
\eeq
 expressed as bilinear of 6 color triplet and 6 
anti-triplet  fermionic operators with $i, j = 1, 2, 3$. The full algebra of $E's$ and $U's$
are generators of $U(6)$. For QCD the  unitary evolution, $\exp[ \pm i \hat H]$, in 
the Hilbert space was referred to as the {\tt QCD
Abacus}~\cite{Brower:1997sm} illustrated in  Fig.~\ref{fig:QCDlink}. The magnetic term acts on  3 color fermionic bits
like beads on wires, changing the color flux on each link. 
The only deformation of the algebra is that on {\bf each link
  $\<x,x+\mu\>$}, the forward ($\hat U(x,\mu)$)
and backward ($\hat U^\dag(x,\mu)$)  gauge matrices  no longer commute ($[\hat U, \hat U^\dag]
\ne 0$) and they are paired with independent left $E_L(x,\mu)$ and right gauge 
$E_R(x,\mu)$.  These no longer obey the identity $E_R = U^\dag  E_L U$. 
However since all link fields  on
two different links still commute, this algebraic deformation is an  irrelevant UV cut-off effect typical
of all lattice field theories. Extended Wilson paths
at long distance will have only $O(a)$
commutators. D-theory conjectures~\cite{Brower:2003vy} that this novel
microstructure   at the cut-off 
under proper implementation can provide a fundamental alternative  to  the Wilson Kogut-Suskind lattice 
gauge theories in the same {\bf universality class} in the continuum limit. 
The precise conditions required to support this
conjecture  of course pose challenging theoretical and  computational  problems.


\section{Quantum Links for U(1) gauge Theory}

To test D-theory operation on existing Quantum Computing hardware, we
now consider the  simplest non-trivial confining gauge theory: the
2+1 Abelian U(1) theory on a triangular lattice.  The compact $U(1)$   group manifold itself is a  circle $S^1$, and the gauge algebra is realized as 
\beq
E_L= -E_R =- i \partial_\theta  \quad , \quad  U= \exp(i\theta) \; .
\eeq
The single link compact manifold, $L^2(S^1)$, is infinite dimensional
but it is a discrete
set of Fourier modes in the flux basis, $\ket q$,  enumerated by integer or half integer  wave number $q$. In this Hilbert space we have that  the action of $E_L, U$ is given by
$E_L\ket q = q \ket q , \quad U \ket q = \ket{q+1}
$. 
It is easy to check that $[E_L,U]= U$ and that this representation
is irreducible.  When we truncate the charge basis, $U$ and $U^\dag$ needs to
annihilate the largest and smallest  charge states respectively.  

D-theory replaces this manifold by the simplest truncation with a  
fermion bilinear representation,  $\hat U  = b^\dag a$,  on a two  Qubit state. However
since Fermion number, $\hat N = a^\dag a + b^\dag b$, is conserved on each link,  
we may project onto the two states with $\hat N = a^\dag a + b^\dag b = 1$ and
faithfully represent the D-theory algebra in terms of Pauli matrices, 
\beq
\hat U = \sigma^+  \quad ,\quad \hat U^\dag  =  \sigma^- \quad, \quad \hat
E_L = - \hat E_R = \sigma^z\; ,
\eeq
in a single Qubit space. Note in this Abelian example both left and right gauge generators at
the ends of each link are the same up to  a sign.
Relative to the full $U(1)$ manifold, now  the flux is  restricted to $\pm 1/2$ on each
link.  

We need to extend this setup to allow for larger values of the flux. 
The angular momentum algebra can serve as a starting point for a representation of the equations $[E_L,U]= U$, $[E_L,U^\dagger]=-U^\dagger$, with $E=L_z$ and $U\propto L^+$, so that the electric field squared is not just a constant like in the spin $1/2$ representation.
We can get higher angular momentum  representations by addition of angular momentum.  This way we write the $E,U$ in terms of sums of spin $1/2$ representations.
This decomposition by addition of angular momentum, allows us to define small kernels 
for   existing NISQ  hardware.

\section{Hamiltonian for 2 +1  $U(1)$ gauge theory}

In D-theory higher flux as we approach the continuum
is built up by  coherent states either in an {\bf extra dimension}  and/or at {\bf larger spatial distances}.
The dynamics of this coherence is an example of the typical
physical mechanism requirement at a second order phase point in order to 
take the continuum limit. If achieved,  universality requires
a similar phase boundary familiar  in the classical example of the Ising vs the $\phi^4$ Hamiltonian
dynamics. 

We consider the 2 + 1 d  $U(1)$ gauge theory Hamiltonian
on a 2d spatial lattice. Beginning with a single plaquette
with $\pm 1/2$ quantum links we build up
high flux by  stacking them in an extra dimension
and extending them  on finite spatial volume. We 
being with the single spatial plaquette as the simplest
toy model. 

\subsection{Single Gauge Plaquette Model}
\begin{figure}[h]
\begin{center}
  \centering
  \subcaptionbox{Electric}
      {\includegraphics[width=.25\linewidth]{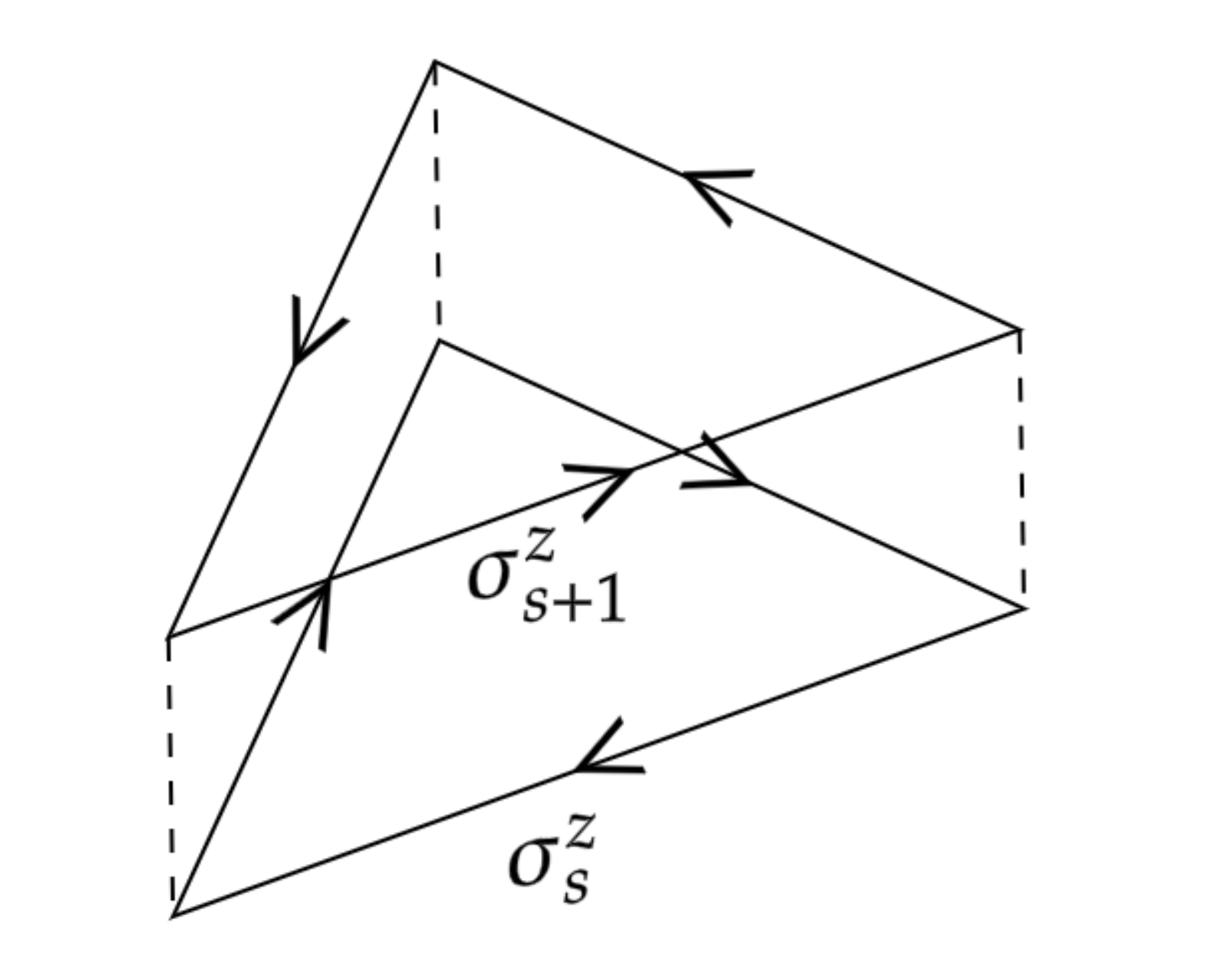}}
  \subcaptionbox{XY Coupling}
      {\includegraphics[width=.3\textwidth]{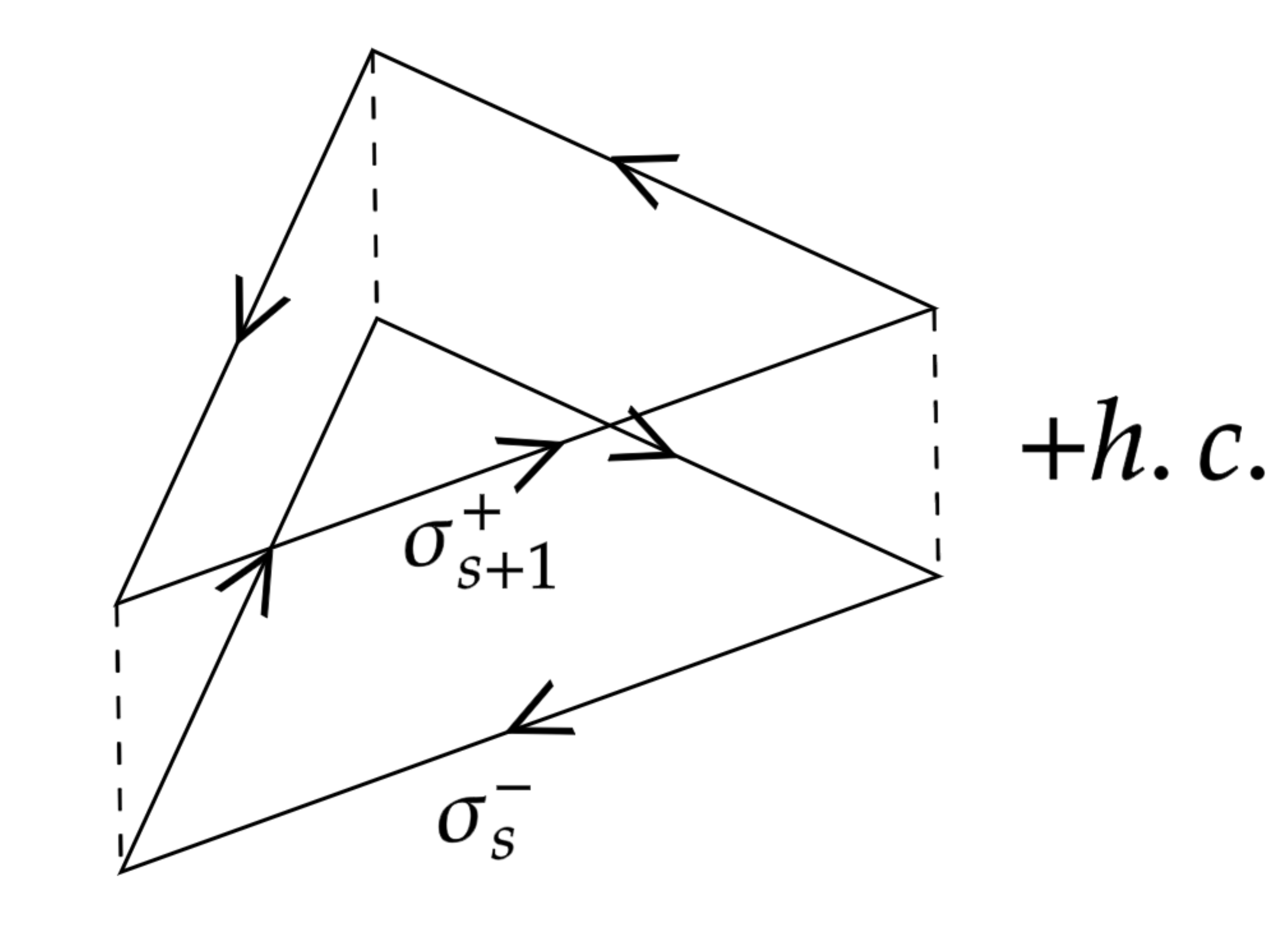}}
  \subcaptionbox{Plaquette}
      {\includegraphics[width=.3\textwidth]{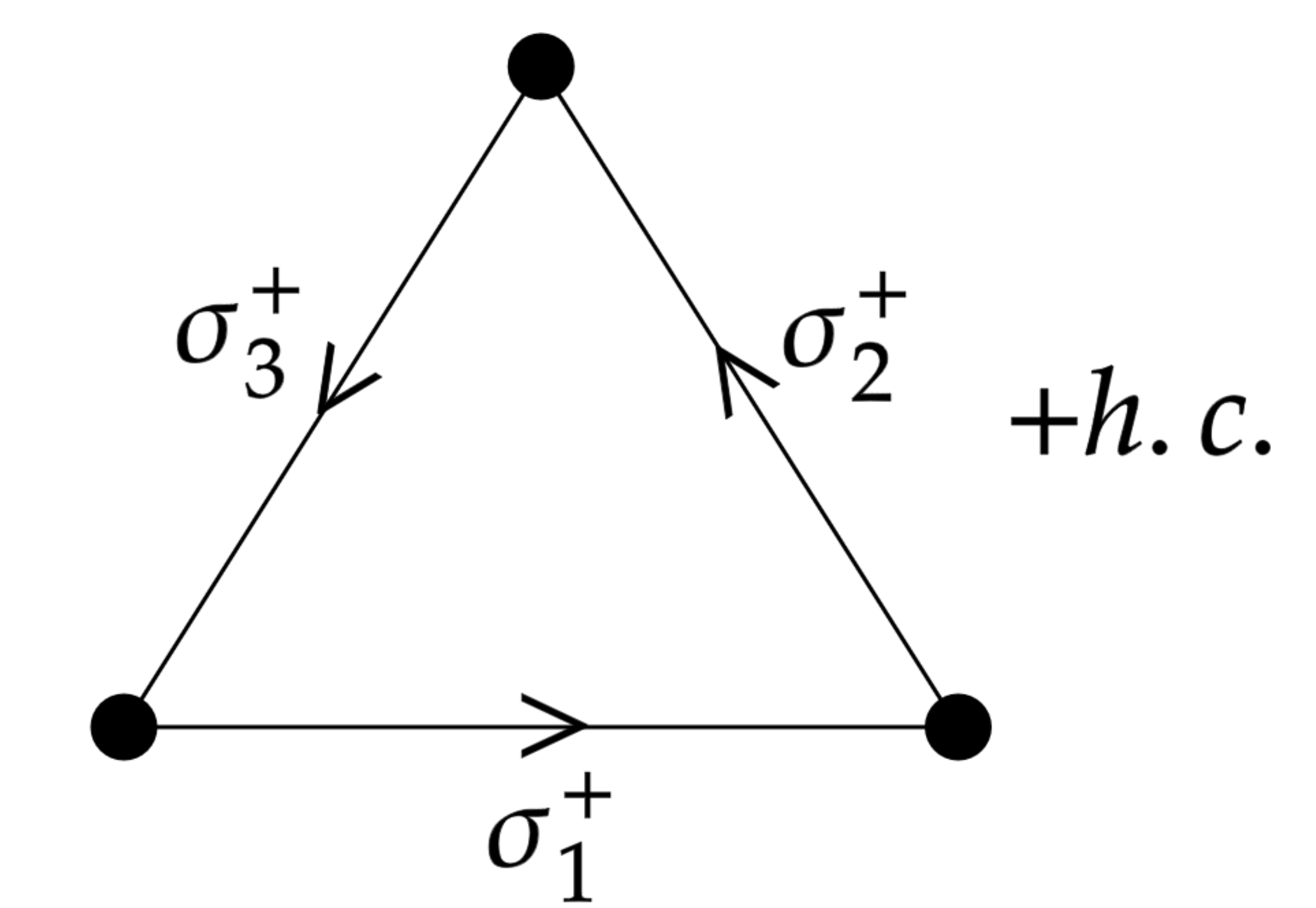}}
\end{center}
\caption{\label{fig:triangle-lattice} Visualization of each term of the Hamiltonian of the single triangle plaquette model with two layers.}
\end{figure}

The Hamiltonian for a single triangular plaquette stacked up
in an extra dimension enumerated by $s$  with periodic boundary conditions is 
\begin{equation}
\begin{split}
    \hat{H} = \sum_s  \bigg[\quad\frac{g^2}{2}\sum^3_{j =1} (\sigma^z_{j, s} + \sigma^z_{j, s+1})^2 
    &+ \frac{\alpha}{2g^2}\sum^3_{j =1} (\sigma^+_{j, s}\;\sigma^-_{j, s+1} + \sigma^-_{j, s}\;\sigma^+_{j, s+1})\\
    & - \frac{1}{2g^2}(\sigma^+_{1,s}\;\sigma^+_{2,s}\;\sigma^+_{3,s} + \sigma^-_{1,s}\;\sigma^-_{2,s}\;\sigma^-_{3,s} )\quad \bigg] \; .
\end{split}
\end{equation}
The lattice structure of this spin-basis Hamiltonian is as in Fig. \ref{fig:triangle-lattice} with triangle links, each of which can be represented as one Qubit spatial link, stacked in the direction of the extra dimension. 
The first term corresponds to the electric term $\hat{E}^2$ of \eqref{eq:Hhat}, and third term  
to the plaquette term of \eqref{eq:Hhat}  represents the magnetic $\hat{B}^2$ part of the energy. 
The positive and negative flux  is created by $\sigma^+$  and  $\sigma^-$ respectively. 
Notice that on a single link $\hat E^2$ is constant, so we need at least one nearest neighbor to separate states based on differences of total electric flux.

The second term can be rewritten as a sum of XX and YY interactions: $\sigma^+_{j, s}\;\sigma^-_{j, s+1} + \sigma^-_{j, s}\;\sigma^+_{j, s+1} = \frac{1}{2}(\sigma^x_{j, s}\;\sigma^x_{j, s+1} + \sigma^y_{j, s}\;\sigma^y_{j, s+1})$. It is an anti-ferromagnetic  nearest neighbor XY spin 1/2 chain for each link.  Together with the first term, it composes a 1D spin-1/2 anisotropic XXZ model.  
There are no gauge variables on links into the extra direction so that the gauge theory still has only
the proper 2d spatial gauge rotations at each site. For example for
a single $\triangle(1,2,3)$ plaquette model the  generators are given by 
\beq
E_{ij} = \sum_s (\sigma_{i, s}^z - \sigma_{j, s}^z) \;
\eeq
where $(i, j) = (1, 2), (2, 3), (3, 1)$. 
Given the periodic conditions and the commutation relations $[\sigma^z_j,\sigma^\pm_i]=\pm\delta_{ij}\sigma^\pm$, we may easily see that all three gauge operators satisfy $[E_{ij}, \hat{H}_P]=0$. The integer flux values for
$E_{ij} $ at each site represent exact conserved Gauss' law  sectors. 

\begin{figure}[h]
    \centering
    \includegraphics[width=.4\textwidth]{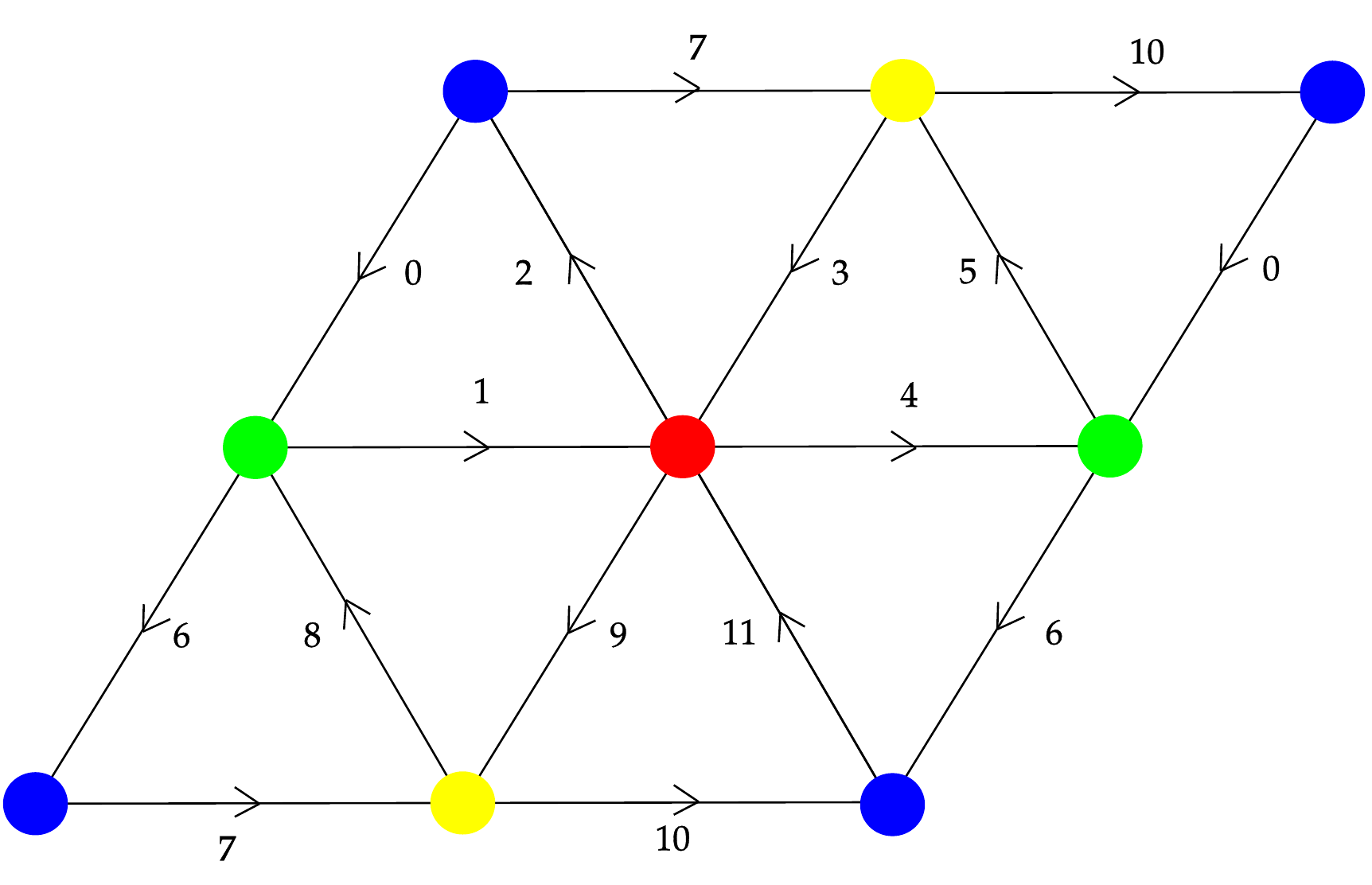}
    \qquad
    \includegraphics[width=.4\textwidth]{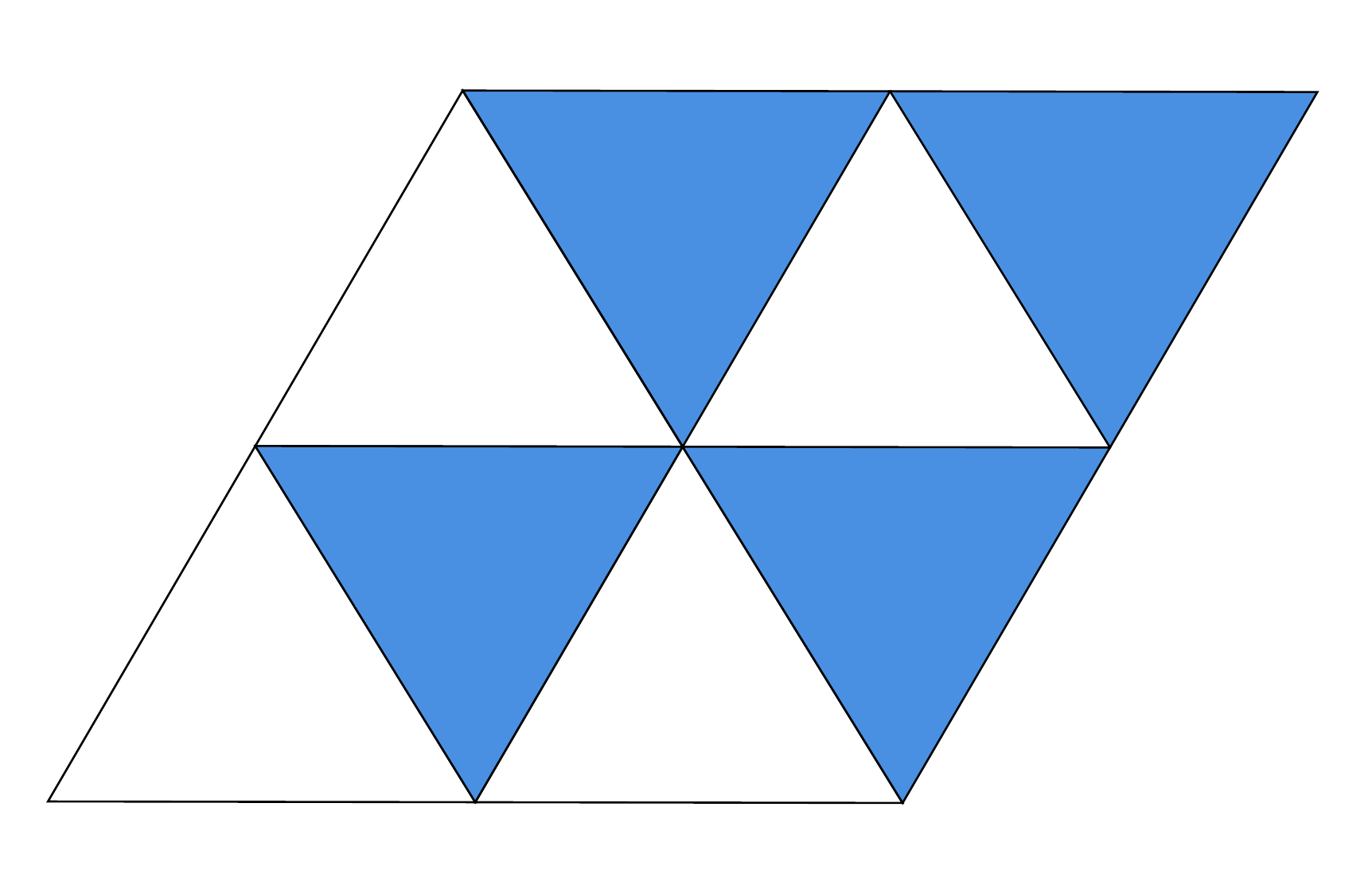}
    \caption{Example of  triangulations of two dimensional spaces for $2\times 2$ torus.
     Tessellation of $2\times 2$ torus. On the left the lattice has 12 independent links (labeled with numbers), 4 independent vertices (labeled with colors) and 8 independent faces. With two layes this is 24 Qubit Hamiltonian On the right we  color the lattice into two sets of face  as a bipartite dual hexagonal lattice. }
    \label{fig:multiple-triangles}
\end{figure}

Now, we can expand this model to finite volume (areas  in the two spatial dimensions)  with multiple triangles
for very small systems with periodic boundary conditions giving a 2d torus. For example
in Fig. \ref{fig:multiple-triangles} is a lattice of eight plaquettes triangulating  a space for a  $2\times 2$ torus, but we may easily extend this to larger than $2\times 2$. The triangle plaquette tiling forms a bipartite dual hexagonal lattice
in Fig. \ref{fig:multiple-triangles}(right), where the Pauli terms of the Hamiltonian associated with one fixed color commute with each other.
This leads naturally to a term Trotter expansion where one can parallelize 
the evolution on a quantum computing algorithm  with 
gauge invariant kernels on plaquettes.
These are implemented next as quantum circuits for single 2 layer single plaquette model. 
 
\subsection{Gauge Invariant Trotter Decomposition and Quantum Circuits}

\begin{figure}[h]
\centering
\subcaptionbox{Electric ($e^{-i\theta\sigma^z\otimes\sigma^z}$) \label{fig:qc-E}}
    {\includegraphics[width=.3\textwidth]{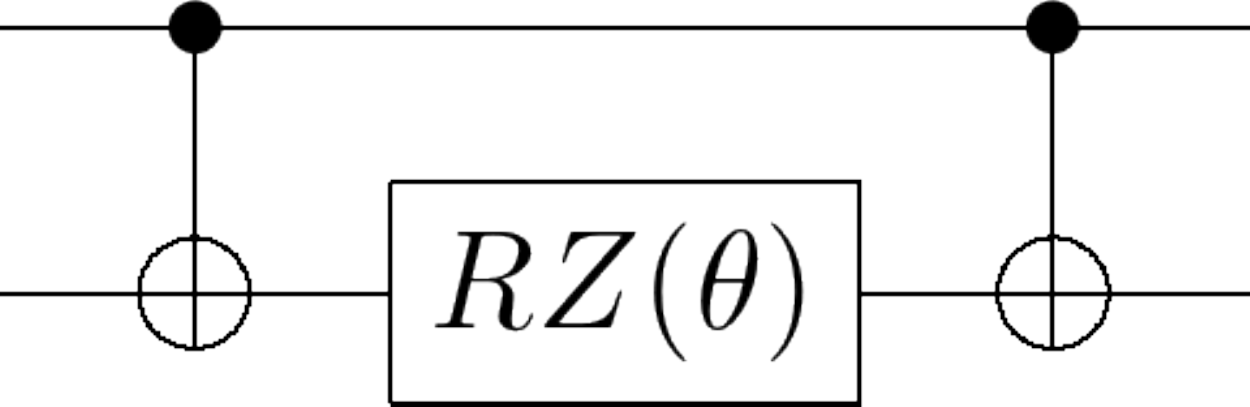}}
\qquad \qquad
\subcaptionbox{XY term ($e^{-i\theta(\sigma^+\otimes\sigma^- + h.c.)}$) \label{fig:qc-XY}}
    {\includegraphics[width=.3\textwidth]{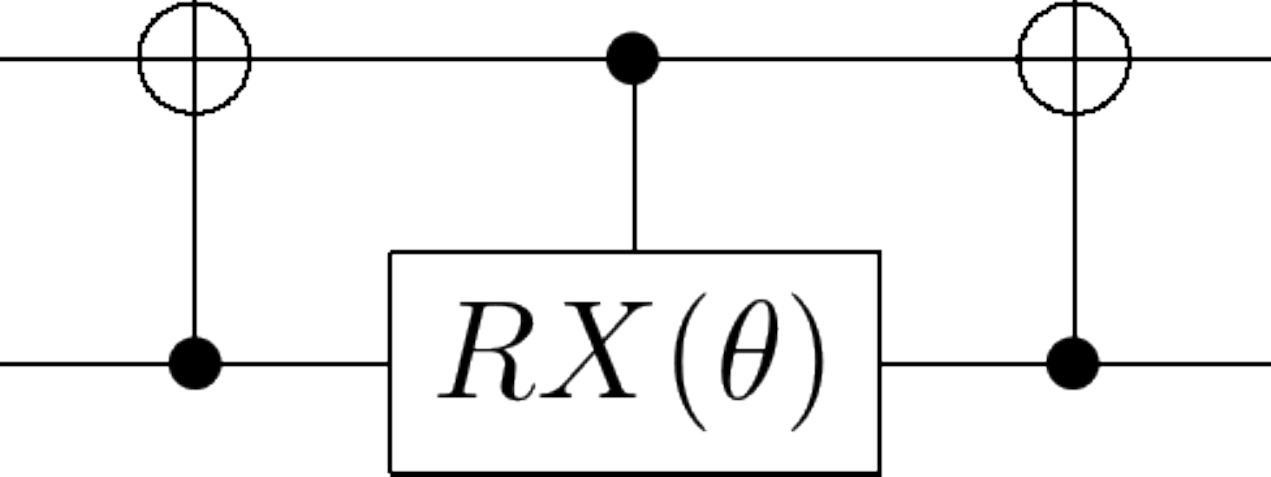}}
\quad
\subcaptionbox{Plaquette $\triangle$\label{fig:qc-B}}
    {\includegraphics[width=.4\textwidth]{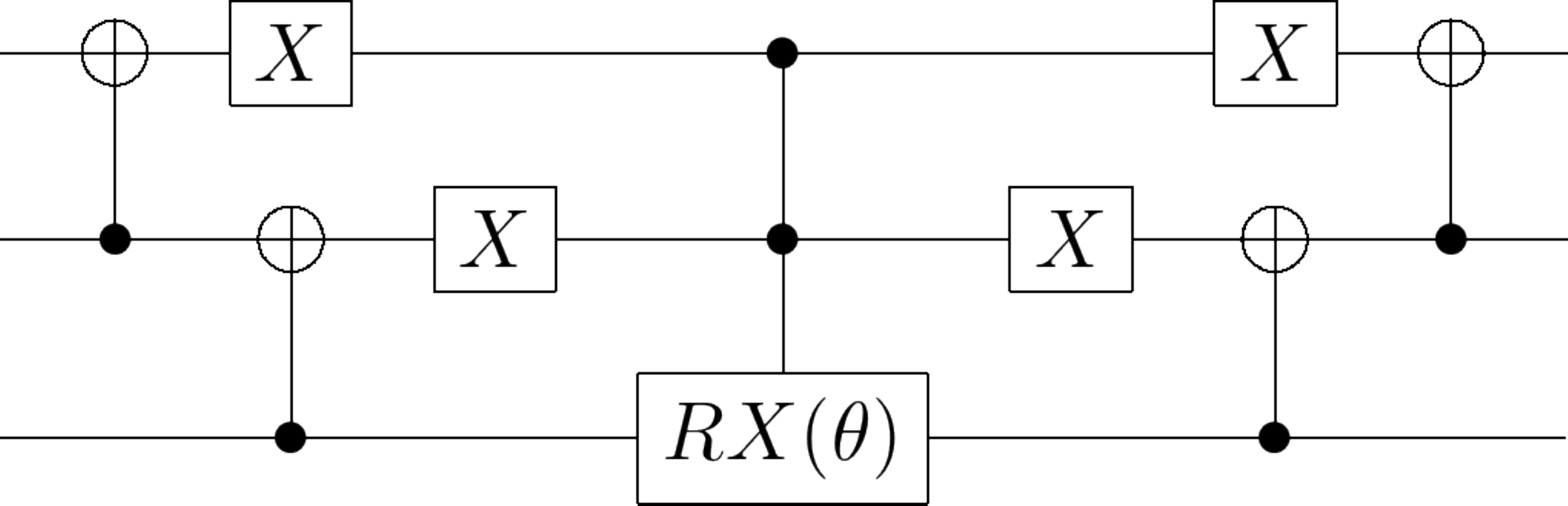}}
\quad
\subcaptionbox{Overall step \label{fig:qc-all}}
    {\includegraphics[width=.45\textwidth]{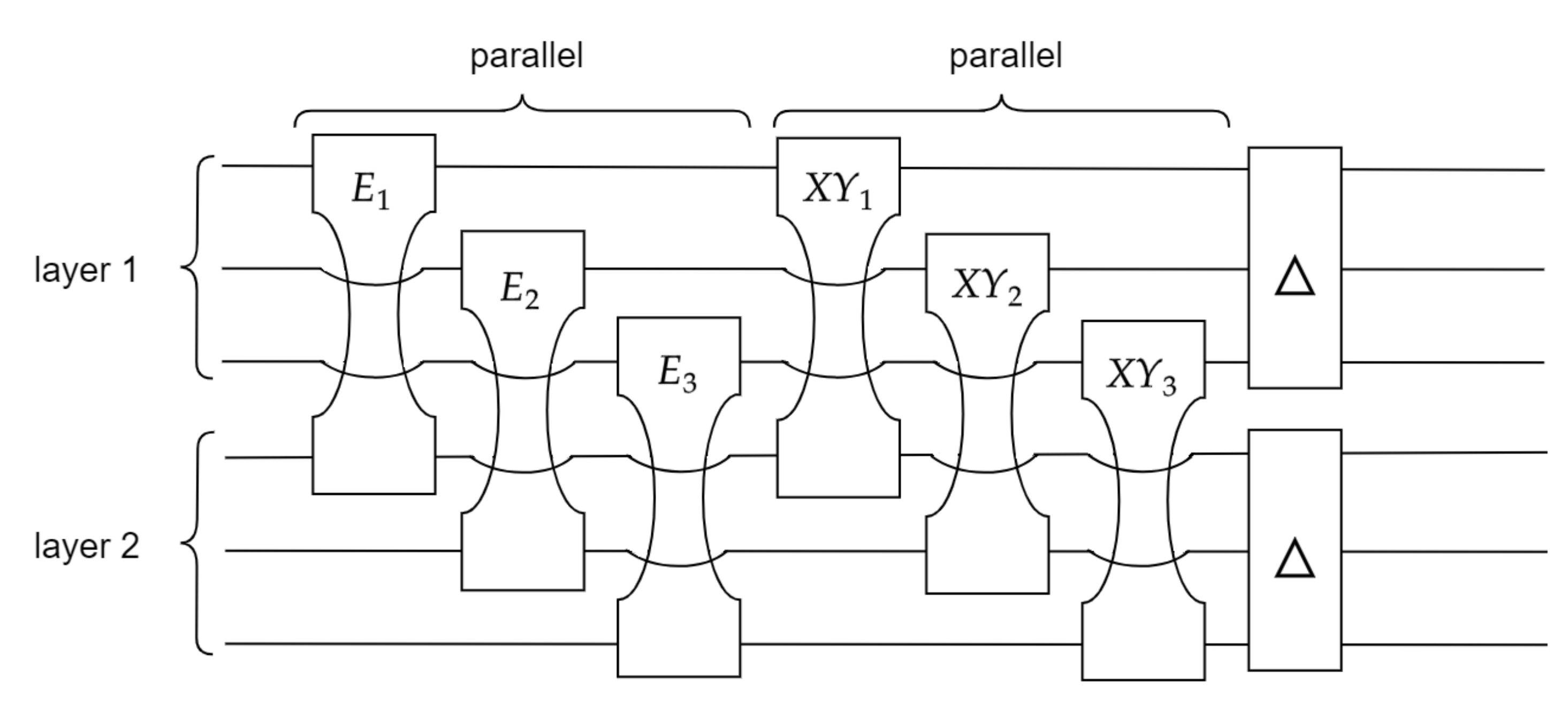}}
\caption{The quantum circuit representing each term of the Hamiltonian of the single plaquette model (a)-(c), and the circuit overview for the entire one single Trotter step (d).} 
\label{fig:trotter-step}
\end{figure}

\begin{figure}[h]
    \centering
    \includegraphics[width=0.5\linewidth]{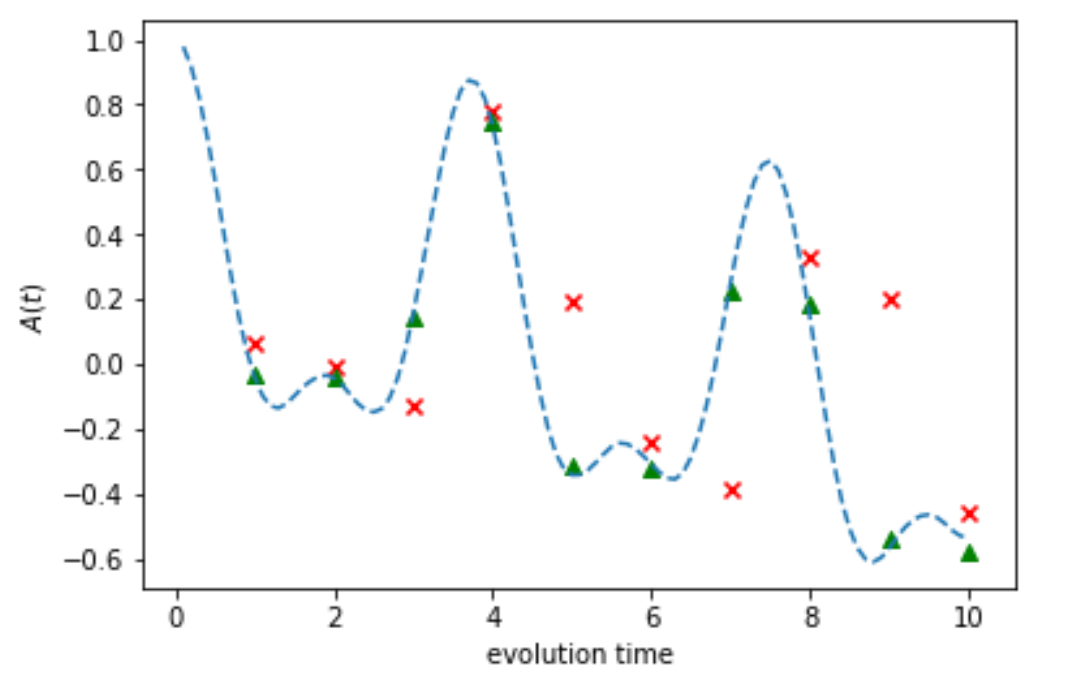}
    \caption{Persistence amplitude $A(t)=\bra{\phi}e^{-i\hat{H}t}\ket{\phi}$ for a single triangle model  with $g = 0.75$ and $\alpha = 1$ from $t=1$ to $t=10$. The state $\ket{\phi}$ is
    initialized in the $\ket +$ state for all Qubits.
    The blue dotted line represents the exact values of this inner product, the red and green points represent the Trotter result with a time interval $\Delta t = 0.5$ and $\Delta t = 0.125$, respectively.}
    \label{fig:trotter_result}
\end{figure}

In order to simulate the time evolution $e^{-i\hat{H}t}\ket{\psi}$ of the plaquette model on a digital quantum computer, the interactions between the links are represented as entanglements between the corresponding Qubits. 
Each of the exponentials of $\hat{E}^2$, coupling, and $\hat{B}^2$ terms can be represented as a quantum circuit as Fig. \ref{fig:qc-E}, \ref{fig:qc-XY}, and \ref{fig:qc-B} respectively. The coupling constant information and the evolution time $t$ in encoded in the rotation angle of the $RZ$ and $RX$ gates. The $XX$ and $YY$ terms commute with each other and together are gauge invariant. The other two are gauge invariant on their own.
The separate non-commuting terms in $\hat{H}$ can be approximated \textit{via} the Trotter decomposition exactly maintaining gauge invariance and unitarity.
Introducing  small time intervals $\Delta t = t/n$, the Trotter
evolution for  $n$   steps is
\begin{equation}
    e^{-i\hat{H}t}\ket{\psi} = \bigg(e^{-i\hat{H}_E\Delta t}e^{-i\frac{\alpha}{2g^2}\sum^3_{j =1} (\sigma^+_{j, s}\;\sigma^-_{j, s+1} + \sigma^-_{j, s}\;\sigma^+_{j, s+1})\Delta t}e^{-i\hat{H}_B\Delta t}\bigg)^n\ket{\psi} \; .
\end{equation}
For the single plaquette model, one Trotter step is  illustrated in Fig. \ref{fig:qc-all}. 
Since different Qubit Pauli matrices commute with each other, the multiple terms
for $\hat{E}^2$, $\hat{B}^2$, and the inter layer XY coupling may be implemented in parallel on a quantum circuit. 
Fig. \ref{fig:trotter_result} demonstrates an example of the time evolution for the single plaquette model Hamiltonian with Trotter decomposition with two different time intervals $\Delta t = 0.5$ and  $\Delta t = 0.125$, compared with the exact computation.

For the multiple plaquette model, we may simply duplicate and repeat this kernel, adding only one additional layer for the two groups of the plaquette terms with different colors which do not commute with each other.

\section{Future Directions}

We have discussed a possible implementation of the $U(1)$ gauge theory in $2+1$ dimensions
on a quantum computer made of Qubits. 
The small kernels identified in this paper are typical of kernels encountered in a variety of 
theories. Investigating alternative Qubit circuit implementations and testing them in early
 Noisy Intermediate-Scale Quantum (NISQ) platforms is useful. This is a process of both refining 
 code kernels and evaluating quantum hardware systems. 
 
 D-theory for $d =4$ gauge and $d =2$ chiral spin theories with asymptotic freedom
 scaling have a clear path to the continuum due to the  gapless phase in $d+1$.  For the Abelian case, a plausible  route
 to the coherent formation of large flux at physical scales  is 
 less well understood.  Perhaps a promising  avenue of investigation 
 of this two state triangular lattice Abelian quantum link model is to look at its duality to a spin model on a hexagonal lattice following
 arguments similar to the classic  duality of the  3d of  Ising model and Wegner’s 3d $Z_2$  gauge theories
 on a square lattice. 
 
Finally there are many interesting small quantum lattice prototypes.  For example
 in addition to toroidal geometries with periodic boundary condition, one may consider  spherical or de-Sitter lattices starting
 with the  sequence  of platonic solids. One can  even  consider small hyperbolic or Anti-de-Sitter lattices generated by the triangle group~\cite{Brower:2019kyh}.
 In conclusion even within the limited context of $U(1)/Z_2$ quantum
 links in the $2+1$ setups, there is a rich theory landscape for testing methods and  confronting fundamental problems  in quantum computing.

 \section*{Acknowledgements}
We are grateful to Cameron V. Cogburn for fruitful discussions. This work was supported in part by the U.S. Department of Energy (DOE) under Award No. DE-SC0019139.

\end{document}